\newdimen\mathindent
\renewcommand{\qquad}{\hspace*{25pt}}
\newcommand{\eref}[1]{(\ref{#1})}
\def\uno{1\!\! 1}

\newcommand{\eqll}[2]{\begin{equation}{#1}\label{#2}\end{equation}}

\def\integer{{\mathbb{Z}}}
\def\real{{\mathbb{R}}}

\documentstyle[11pt,epsf,amssymb]{article}
\begin{document}

\title{Allowed and observable phases in two-Higgs-doublet Standard Models}
\author{G. Sartori and G. Valente}

\author{G Sartori and G Valente\\
\small Dipartimento di Fisica,Universit\`a di Padova \\
and INFN, Sezione di Padova\\
\small via Marzolo 8, I--35131 Padova, Italy \\
\footnotesize (e-mail: gfsartori@padova.infn.it,
valente@padova.infn.it)}
\date{}

\maketitle

\begin{abstract}
In Quantum Field Theory models of electro-weak interactions with
spontaneously broken gauge invariance, renormalizability limits to
four the degree of the Higgs potential, whose minima determine the
possible vacuum states in tree approximation. Through the
discussion of some simple variants of the Standard Model with two
Higgs doublets, we show that, in some cases, the technical limit
imposed by renormalizability can prevent the observability of some
phases of the system, that would be otherwise allowed by the
symmetry of the Higgs potential. An extension of the scalar sector
through suitable SU$_2$  singlet particle fields
can resolve this {\em unnatural} limitation.

\end{abstract}

\section{Introduction}
\label{intro}

In the Standard Model (SM) of Electro-Weak (EW) interactions
\cite{GWS}, although the gauge boson and fermion structure has
been accurately tested, experimental information about the Higgs
sector (HS) is still very weak. Serious motivations are well known
for the extension of the scalar sector; among them, we just recall
supersymmetry (SUSY) and baryogenesis at the EW scale (see, for
instance, \cite{1,2}  and references therein). So far, various
extensions of the SM have been devised: the Minimal SUSY SM, the
SM plus an extra Higgs doublet, the MSSM plus a Higgs singlet, the
left--right symmetric model, the SM plus a complex singlet Higgs
(see the introduction to \cite{3} and references therein);
recently, even a partly supersymmetric SM has been conceived
\cite{4}.

Our paper is devoted to the discussion of one of the most widely
accepted paradigms of quantum field theory, namely the
renormalizability constraint imposed on the Higgs potential. We
shall show how symmetry may be used as a guideline to the
construction of the scalar sector of the theory.

In QFT models with broken gauge invariance, the symmetry group $G$
and the transformation properties of the scalar fields under $G$
(hereafter simply, and improperly, referred to as {\em symmetry}
of the model) determine the isotropy subgroup chain, that is all
the phases that are permitted by the symmetry (hereafter called
{\em allowed phases}). Not all the allowed phases need be
observable in the evolution of the Universe. In fact, only the
allowed phases which correspond to a scalar field configuration
determining a local minimum of the Higgs potential are
``dynamically attainable"; moreover, perturbatively unstable
phases have zero probability to be seen. So, we shall call {\em
observable} only the phases that are determined by stable (against
small perturbations of the coefficients) local minima of the Higgs
potential. If the Higgs potential is an invariant polynomial of
sufficiently high degree, with arbitrary coefficients ({\em
phenomenological parameters}), all the allowed phases turn out to
be dynamically attainable \cite{AS}. However, to ensure the
renormalizability of the theory, the degree of the Higgs potential
has to be $\leq 4$ and, in some models (hereafter called {\em
incomplete}), this bound prevents some allowed phases to be
observable.

Our point of view is that renormalizability, which actually has to
be considered a ``technical'' assumption required to assure a
consistent and significant perturbative solution of the theory,
should not limit the implications of the basic symmetry of the
formalism used to describe the system at the classical level
\cite{Gufan}. All the {\em
allowed} phases have to be {\em observable}. This attitude may
have important consequences in the study of EW phase transitions,
in the sense that the allowed phases have to be thought, in
principle, as possible phases in the evolution of the Universe
\cite{5}.

Below, we shall illustrate our claims in some examples obtained
from simple extensions of the Higgs sector of the SM. In
particular we shall show that, while in the SM all the allowed
phases are observable, in models with two Higgs doublets, if the
usual additional discrete symmetries are added to avoid flavour
changing neutral current (FCNC) effects, this is true only if the
Higgs potential is a polynomial of sufficiently high degree,
greater than 4. We shall also show that all the allowed phases of
a renormalizable two Higgs model can be made observable, if
$\mathrm{SU}_2$ singlet ``composite" scalar fields, with suitable transformation
properties under the discrete symmetries are included.

\section{The geometrical invariant theory approach to spontaneous symmetry breaking}

A general approach to the determination of all the allowed phases
has been proposed in \cite{AS}. Let us briefly recall the basic
elements, which will be essential for the derivation of our
results.

The set of real scalar fields of the model will be denoted by
$\phi$, to be thought of as a vector in $\real^{n}$ (vector order
parameter), transforming according to a real orthogonal
representation of the gauge group. We shall denote by $G$ the
representative real linear group. The Higgs potential $V_a(\phi)$ is a
$G$-invariant real polynomial function of $\phi$ with coefficients
$a_i$. The points of stable local minimum of $V_a(\phi)$ determine
the stable phases of the system. Owing to $G$-invariance, the
Higgs potential is a constant along each $G$-orbit, so, each local
minimum is degenerate along a whole $G$-orbit, whose points define
equivalent vacua. Since the isotropy subgroups of $G$ at points of
the same $G$-orbit are conjugate in $G$, only the conjugacy class
in $G$ of isotropy subgroups of $G$ at the points of the orbit of
equivalent minima, {\em i.e.} the {\em orbit-type} of the orbit, is
physically relevant, and defines the symmetry of the associated
phase.

The set of all $G$-orbits, endowed with the quotient topology and
differentiable structure, forms the {\em orbit space},
$\real^n/G$, of $G$ and the subset of all the $G$-orbits with the
same orbit-type forms a {\em (symmetry) stratum} of $\real^n/G$.
At the classical level, phase transitions take place when,
by varying the values of the
$a$'s, the minimum of $V_a(\phi)$ is shifted to an orbit lying on
a different stratum \cite{isos}.

If $V_a(\phi)$ is a polynomial in $\phi$, of sufficiently high
degree, by varying the $a$'s, its absolute minimum can be shifted
to any stratum of $\real^n/G$. So, {\em the strata are in a
one-to-one correspondence with the allowed phases}. On the
contrary, extra restrictions on the form of the Higgs potential,
not coming from G-symmetry requirements (e.g., the assumption that
it is a fourth-degree polynomial), can prevent its local minima
from sitting in particular strata as perturbatively stable minima
and make, consequently, the corresponding allowed phases
dynamically unattainable.

Being constant along each $G$-orbit, the Higgs potential can be
conveniently thought of as a function defined in the orbit space
$\real^n/G$ of $G$. This fact can be formalized using some basic results of
invariant theory. In fact, every $G$-invariant polynomial function
$F(\phi)$ can be built as a real polynomial function $\widehat
F(p)$ of a {\em finite} set, $\{p_1(\phi), \dots ,p_q(\phi)\}$, of
basic homogeneous polynomial invariants ({\em minimal integrity basis (MIB) of the ring of
$G$-invariant polynomials):}

\eqll{F(\phi)=\widehat F(p(\phi)),\quad \phi\in\real^n}{F} and the
range $p(\real^n)$ of the {\em orbit map}, $\phi \mapsto p(\phi) =
(p_1(\phi), \dots ,p_q(\phi))\in \real^q$ yields a diffeomorphic
realization of the orbit space of $G$, as a connected
semi-algebraic set, {\em i.e.}, as a subset of $\real^q$
determined by algebraic equations and inequalities. Thus, the
elements of an integrity basis can be conveniently used to
parametrize the points of $p(\real^n)$ that, hereafter, will be
identified with the orbit space $\real^n/G$.

The elements of a minimal integrity basis $\{p\}$ need not, for
general compact groups, be algebraically independent (possible
algebraic relations among the elements of a MIB are called {\em
syzygies}). The number $q_0$ of algebraically independent elements
in $\{p\}$ is $n$ minus the dimension as a manifold of a generic
{\em (principal)} orbit of $G$. The semialgebraic set $p(\real^n)$
has been shown to be formed by the points $p\in \real^q$,
satisfying the following conditions i) and ii) \cite{AS}:

\begin{itemize}
\item[i)] $p$ lies on the surface, $Z$, defined by the syzygies;
\item[ii)] the following $q\times q$ matrix $\widehat P(p)$ is
positive semi-definite and has rank $\le q_0$ at $p$:

\begin{equation}
\widehat P_{ab}(p(\phi)) = \sum_{j=1}^n\partial_j
p_a(\phi)\,\partial_j p_b(\phi),\quad \forall \phi\in \real^n.
\end{equation}

\end{itemize}

Like all semi-algebraic sets, the orbit space of $G$ presents a
natural {\em stratification}. It can, in fact,  be considered as
the disjoint union of a {\em finite number} of connected
semi-algebraic subsets of decreasing dimensions ({\em primary
strata}), each primary stratum being a manifold lying in the
border of  higher dimensional ones, but for the highest
dimensional stratum, which is unique ({\em principal stratum}).
The primary strata are the connected components of the {\em
symmetry strata}. The symmetries of two bordering strata are
related by a group--subgroup relation and the orbit-type of the
lower dimensional stratum is larger. If the only $G$-invariant
point is the origin of $\real^n$, in $\real^n/G$ there is only one
0-dimensional stratum corresponding to the origin of $\real^q$.
All the other strata have at least dimension 1, since the isotropy
subgroups of $G$ at the points $x\in\real^n$ and $\lambda x$,
$\lambda\in\real$, are equal and, therefore, the points
$(p_1,\dots ,p_q)$ and $(\lambda^{d_1}p_1,\dots
,\lambda^{d_q}p_q)$ ($d_i$ the degree of the basic invariant
$p_i(x)$) sit on the same stratum. This fact, added to the
homogeneity of the basic invariants and of the relations defining
the strata, shows also that a complete information on the
structure of the orbit space and its stratification can be
obtained from its intersection with a hyperplane, which is the
image in $\real^n/G$ of the unit sphere \cite{nota} of $\real^n$.

By defining, according to \eref{F},

\begin{equation}
\widehat V_a(p(\phi))= V_a(\phi),\qquad \phi\in \real ^n,\label{1}
\end{equation}
the range of $V_a(\phi)$ coincides with the range of the
restriction of $\widehat V_a(p)$ to the the orbit space
$p(\real^n)$ and the local minima of $V_a(\phi)$ can be computed
as the minima of the function $\widehat V_a(p)$ with domain
$p(\real^n)$.

In detail, denoting by $f_\alpha(p)=0$, $\alpha=1,\dots k$ a complete set of
independent equations of the stratum $\widehat \sigma$, the conditions for
a stationary point of the potential at $p\in\widehat\sigma$, can be
conveniently written in the following form:

\eqll{\frac {\partial}{\partial p_i} \left(\widehat
V_a(p)-\sum_{\alpha=1}^k\,\lambda_\alpha\, f_\alpha(p)\right)=0,
\qquad i= 1,\dots ,q}{ext} where the $\lambda_\alpha$'s are
Lagrange multipliers. The stationary point will be a stable local
minimum on the stratum if the Hessian matrix $H(\phi)$ of
$V_a(\phi)$, for any $\phi$ in the orbit of equation $p=p(\phi)$,
is $\ge 0$ and has rank equal to $n$ minus the dimension $\nu$ of
the orbit ($\nu$ equals the number of Goldstone bosons and the
stability condition implies the absence of pseudo-Goldstone bosons
\cite{GePa}). These conditions can be conveniently expressed in
terms of the sums $M_i$ of the (determinants of) the principal
minors $H(\phi)$ in the form $M_i>0$, $i=1,\dots ,n-\nu$. Being
$H(\phi)$ a $G$-tensor of rank 2, the $M_i$'s are $G$-invariant
polynomials in the $\phi_i$'s and can, therefore, be expressed as
polynomials in the elements of the MIB.

In the following, we shall characterize all the allowed and
observable phases and all possible phase transitions between
contiguous phases, for variants of a two Higgs doublets version of
the SM. In particular, for each model we shall determine a minimal
set of basic polynomial invariants of $G$, the geometrical
features of the orbit space, {\em i.e.}, its stratification
 and the orbit-types of its strata.

\section{Model 1} The symmetry group of the Lagrangian is
SU$_2\times$U$_1$ and there are two complex Higgs doublets
$\Phi_1$ and $\Phi_2$ of hypercharge $Y=1$. In this model, natural
flavor conservation is violated by neutral current effects in the
phase (hereafter called $\sigma^{(3)}$) that should correspond to
the present phase of our Universe. So the model is not realistic,
but it provides us a simple example in which renormalizability
does not exclude completeness.

Let us stress that the transformation induced by the element
$\hat{j} = (-\uno_2,\mathrm{e}^{\mathrm{i} \pi Y})
\in$ SU$_2\times$U$_1$ leaves invariant the fields
$\Phi_1$ and $\Phi_2$. So, for our purposes, it will be equivalent,
but simpler, to consider as invariance group
$G= (\mathrm{SU}_2\times \mathrm{U}_1)/\integer_2$, where $\integer_2$
is the group generated by $\hat{j}$.

A convenient choice for a MIB of real polynomial independent
$G$-invariants is the following:
\eqll{p_1=\Phi_1^\dagger\Phi_1 + \Phi_2^\dagger\Phi_2,\ \ p_2 = \Phi_1^\dagger\Phi_1
- \Phi_2^\dagger\Phi_2,\ \ p_3 + \mathrm{i} p_4 = 2\,\Phi_2^\dagger\Phi_1.}{IB1}

The relations defining $p(\real^4)$ and its strata, which are
listed in Table~\ref{T1}, can be obtained from rank and positivity
conditions of the $\widehat P(p)$-matrix \cite{AS} associated to
the MIB defined in Eq.~\eref{IB1}. The non vanishing elements of
$\widehat P(p)$ are $\widehat P_{ii}(p)=4\,p_1,$  $\widehat
P_{1j}(p)=\widehat P_{j1}(p)=4\,p_j,$ for $i=1 \ldots 4$ and $j=2,\ldots,4$.

\begin{table}
\caption{Strata of the orbit space for the symmetry group
$G= (\mathrm{SU}_2\times \mathrm{U}_1)/\integer_2$
of Model 1. The index in
the symbols distinguishing the strata denotes their dimension in
orbit space and the bar denotes topological closure. The group U$_1^{\mathrm e.m.}$
is the set of the elements
$ \left \{ \mathrm{e}^{ \mathrm{i} \theta (T_3+Y/2)} \right \}_{0 \leq \theta < 2 \pi}$.
For each stratum, a field configuration with the same symmetry is supplied {\em (typical point)}.
The fields are represented by their real components:
$\Phi_1^{\mathrm T}=(\phi_1 + \mathrm{i} \phi_2,\phi_3 + \mathrm{i} \phi_4)$,
$\Phi_2^{\mathrm T}=(\phi_5 + \mathrm{i} \phi_6,\phi_7 + \mathrm{i} \phi_8)$
and the $\phi_i$'s are generic non zero values.
\label{T1}}
\begin{center}
\begin{tabular}{llccl}
\hline
Strata   & Defining relations             & Symmetry   & Boundary &  Typical point $\phi$      \\
\hline
$\sigma^{(4)}$ & $p_1>\sqrt{p_2^2+p_3^2+p_4^2}$ & $\{\uno \}$            & ${\overline \sigma}^{(3)}$ & $(\phi_1,0,\phi_3,\phi_4,0,0,\phi_7,0)$\\
$\sigma^{(3)}$ & $p_1=\sqrt{p_2^2+p_3^2+p_4^2}$ & U$_1^{\mathrm e.m.}$ & $\sigma^{(0)}$ &  $(0,0,\phi_3,\phi_4,0,0,\phi_7,0)$\\
$\sigma^{(0)}$ & $p_1= p_2 = p_3 = p_4 = 0$     & $G$ & & $(0,0,0,0,0,0,0,0)$\\
\hline
\end{tabular}
\end{center}
\end{table}

The orbit space is the half-cone bounded by the surface of
equation $p_1=\sqrt{\sum_{i=2}^4 p_i^2}$. The tip of the cone
corresponds to the stratum $\sigma^{(0)}$, and the rest of the surface
to the stratum $\sigma^{(3)}$, while the interior points form
$\sigma^{(4)}$.

The most general fourth-degree polynomial invariant Higgs
potential can be written in the following form:

\begin{equation}
\begin{array}{rcl}
\widehat V(p)&=& \sum_{i,j=1}^4 \,A_{ij}\,p_i\, p_j +
\sum_{i=1}^4\,a_i\, p_i \\
&&\\
&=&  \sum_{i,j=1}^4 \,A_{ij}\,(p_i-\eta_i)(p_j-\eta_j)-
 \sum_{i,j=1}^4 \,A_{ij}\,\eta_i\,\eta_j,
\end{array}\label{V}
\end{equation}
where, to make sure that $\widehat V(p)$ is bounded below, we
assume that the symmetric real matrix $A$ is positive definite and
$\eta_i= - \frac{1}{2} \sum_{j=1}^4 (A^{-1})_{ij}\,a_j$.

In this simple case (convex orbit space), the minima of $\widehat V(p)$ can be easily
determined from elementary geometrical considerations. To this
end, let us first assume $A\propto \uno$ and let us denote by
$C=C^+\cup C^-$  the closed double cone bounded by the
surfaces of equation $\eta_1=\pm\sqrt{\sum_{i=2}^4 \eta_i^2}$.
Then, since the potential is a constant plus the squared distance of
$p$ from $\eta$, for given values of the $\eta_i$'s, there is only
one local minimum of the potential (the absolute minimum) at the
point $p$ of the orbit space which is at shortest distance from
$\eta$. One is left, therefore, with the following possibilities:

\begin{itemize}
\item[i)] the minimum is stable  in $\sigma^{(4)}$, at $p=\eta$, for $\eta$ in the interior of $C^+$;
\item[ii)] the minimum is stable in $\sigma^{(0)}$, at $p=0$, for $\eta$ in the interior of $C^-$;
\item[iii)]  the minimum is stable in $\sigma^{(3)}$, at the point at shortest distance
from $\eta$, for $\eta$ outside $C$;
\item[iii)]  the minimum is unstable in $\sigma^{(3)}\cup \sigma^{(0)}$, at $p=\eta$, for $\eta$ on
the surface of the double cone.
\end{itemize}

For a general $A>0$, the results do not essentially change, since
one can revert to the case $A= \uno$ by means of a convenient
linear transformation of the $p_i$'s, which defines a new
(equivalent) MIB: as a result, $C^+$ and $C^-$ are
 simply rotated and deformed by independent re-scalings along the coordinate axes. So, in the
space of the parameters $(a_1,\dots ,a_4)$, that are independent
linear combinations of the $\eta_i$'s, there are three disjoint
open regions of stability of the three allowed phases associated
to the strata of the orbit space. These regions
are separated by inter-phase boundaries, formed by critical points
where second order phase transitions may start; moreover, first
order phase transitions
cannot take place \cite{cave}.

The evolution of the Universe described by the model can be
represented by a continuous line in the space of the parameters
$(a_1,\dots ,a_4)$. A random path in this space has zero
probability of crossing the origin, so the only observable second
order phase transitions correspond to transitions between the
phases associated to the strata $\sigma^{(4)}$ and $\sigma^{(3)}$ or
$\sigma^{(3)}$ and $\sigma^{(0)}$.

We can conclude that the model just discussed is both {\em renormalizable} and {\em complete}.

\section{Model 2}
The model contains the same set of fields as Model 1, but the
symmetry group of the Lagrangian is assumed to be
SU$_2\times$U$_1\times \{\hat \iota\,,K\}$, where $\hat
\iota$ is the generator of a  reflection group and $K$ is the
generator of $CP$-like transformations:
 $(\Phi_1,\Phi_2) \rightarrow (\Phi_1,-\Phi_2)$ and
 $(\Phi_1,\Phi_2) \rightarrow (\Phi_1^*,\Phi_2^*)$,
respectively. In fact, it is well known that the set of the phases
in the models with two Higgs doublets depends strongly on the
discrete symmetries, which are not entirely fixed by experimental
constraints. To protect the theory from FCNC processes, it is
nowadays commonly accepted the introduction of the $\hat \iota$
symmetry \cite{1,2}. Nevertheless, since the most general $CP$
transformation on a field $X_r$ contains a field--dependent phase
\cite{SWe}, i.e. $X_r \longrightarrow \mathrm{e}^{\mathrm{i}
\theta_r} {X_r}^\ast$, the $CP$ conservation is normally checked
{\em a posteriori}. Our $CP$-like transformations allow an easy
verification of $CP$ conservation. For example, with reference to
Tables~\ref{T2a} and \ref{T2b}  below, it is evident that $CP$ is broken in the stratum
${\tilde S}^{(3)}_3$, while in ${\tilde S}^{(2)}_2$ $CP$ is
conserved: the transformations induced by $\hat{\iota}$
determine the right $CP$-phase $\theta$ for
each field of the theory.

As in model 1, for our purposes it will
be equivalent, but simpler, to consider as a symmetry group of the
model $G= (\mathrm{SU}_2\times \mathrm{U}_1)/\integer_2
\,\times \{\hat \iota\,,\, K\}$.
Under these assumptions, a MIB is the
following:

\begin{equation}
\tilde{p}_1=\Phi_1^\dagger\Phi_1,\;\; \tilde{p}_2=\Phi_2^\dagger\Phi_2,\;\;
\tilde{p}_3=\left(\mathrm{Re}\left[\Phi_2^\dagger\Phi_1\right]\right)^2,\;\;
\tilde{p}_4=\left(\mathrm{Im}\left[\Phi_2^\dagger\Phi_1\right]\right)^2.\label{IB2}
\end{equation}
The strata of $\tilde{p}(\real^8)$ will be
denoted by $\tilde{S}^{(i)}_j$, where $i$ is the dimension of the
stratum and $j$ is an order index. The defining relations are obtained from positivity
and rank conditions of the symmetric matrix $\widehat
P(\tilde{p})$, associated to the MIB of Eq.~\eref{IB2}, whose non
zero (upper triangular) elements are the following: $\widehat
P_{11}(\tilde p)= 4 \tilde{p}_1$, $\widehat P_{22}(\tilde p)= 4
\tilde{p}_2$, $\widehat P_{13}(\tilde p)=\widehat P_{23}(\tilde p)= 4
\tilde{p}_3$, $\widehat P_{14}(\tilde p)= \widehat P_{24}(\tilde p)= 4
\tilde{p}_4$, $\widehat P_{33}(\tilde p)= 4
\tilde{p}_3\,(\tilde{p}_1+\tilde{p}_2)$ and $\widehat
P_{44}(\tilde p)= 4 \tilde{p}_4\,(\tilde{p}_1+\tilde{p}_2)$. The
results can be easily recovered from Tables~\ref{T2a} and \ref{T2b},
 by
identifying
$(p_1,p_2,p_7,p_8)=(\tilde{p}_1,\tilde{p}_2,\tilde{p}_3,\tilde{p}_4)$
and setting to zero all the other $p_i$'s.

\begin{figure}[hbt]
\caption{\footnotesize Section $\Xi$ of the four dimensional orbit space of Model 2
with the hyperplane of equation $\tilde{p}_1+\tilde{p}_2=1$.\label{F1}}
\begin{center}%
\leavevmode
\hbox{%
\epsfxsize=3in%
\epsffile{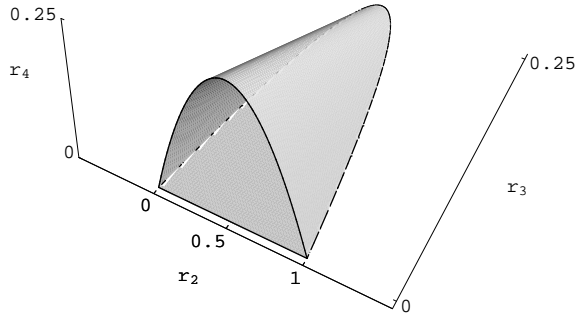}%
}%
\end{center}
\end{figure}

The section $\Xi$ of $\tilde{p}(\real^8) \subset \real^4$ with the hyperplane of
equation $\tilde{p}_1+\tilde{p}_2=1$ is the three dimensional closed
solid (semialgebraic set) drawn in
Figure~\ref{F1}. The full orbit space is the four dimensional
connected semi-algebraic set formed by
the points $\tilde{p}= (\tilde{p}_1,\tilde{p}_2,\tilde{p}_3,\tilde{p}_4)=  \Pi^{-1} (r)$,
$ r \in \Xi$, where $\Pi^{-1}$ is the inverse
projection defined as follows:
\begin{equation} \label{proinv}
\begin{array}{lccl}
\Pi^{-1}: & \real^3 \supset \Xi & \longrightarrow & \real^4\\
  & (r_2,r_3,r_4) & \mapsto &
 \left( \lambda (1-r_2),\lambda r_2,\lambda^2 r_3,\lambda^2 r_4 \right)\,,\;\;\;\;\lambda \geq 0\,.
\end{array}
\end{equation}

% $(\tilde{p}_1,\tilde{p}_2,\tilde{p}_3,\tilde{p}_4)=
% \left( \lambda (1-r_2),\lambda r_2,\lambda^2 r_3,\lambda^2 r_4\right)$,
% with $\lambda \geq 0$ and
%$(r_2,r_3,r_4)\in\Xi$.

%%%%%%%%%%%%%%%%%%%%%%%%%%%%%%%%%%%%%%%%%%
\begin{table}
\caption{\label{T2a} \footnotesize Orbit space characterization of strata $S$
of Model 3 and strata $\tilde{S}$ of Model 2.
The $s_i$ are given in Eq.~(\ref{Sy}), $ q =p_1 p_2 - p_7 - p_8 $,
The
relations defining the $\tilde{S}$'s are recovered setting
$(p_1,p_2,p_7,p_8)=(\tilde{p}_1,\tilde{p}_2,\tilde{p}_3,\tilde{p}_4)$
and $p_i=0$ otherwise. Neighbouring strata are given, so that possible second order
phase transitions can  be easily identified.}
\begin{center}
\begin{tabular}{lll}
\hline
 Stratum   & Defining relations                   &  Boundary \\
 \hline
 $S^{(7)} \equiv \tilde{S}^{(4)} $   & $s_i = 0 \leq q\,,i=1,\ldots,6\,; p_j \geq 0, \; j= 3,4,7,8 $ & ${\overline S^{(5)}}\,;\;{\overline S_i^{(4)}}\,,\;i=1,2,3$\\
                        & $p_1\,,p_2>0\,;\;\; (p_1+p_2)p_4+p_7+p_8>0$ &\\
           & $p_3\,(p_1+p_2)+q >0$ &\\
            & $p_4(p_1+p_2)+p_8+(q p_3-p_5^2)>0$ &\\
             & $p_7(p_1+p_2)+(q p_3-p_5^2)>0$ &\\
  $S^{(5)}\equiv \tilde{S}^{(3)}_1$   & $ s_1 = p_5^2-q\,p_3=0= p_i,\,\ i=7,9,10$ & ${\overline S_2^{(2)}}\,;{\overline S_i^{(3)}},\; i=1,2$ \\
&$ q\,,\;p_i \geq 0\,,\;i=1,2,3,4,8 $&   \\
&$p_1+p_2>0\,,\;(p_1+p_2)p_4+p_8 >0 $&   \\
&$q+(p_1+p_2)p_3 >0 $&   \\
 $S^{(4)}_1\equiv \tilde{S}^{(3)}_2$ & $ p_5^2 - p_3\,q = 0 = p_i,\ i=4,6,8,9,10$     & ${\overline S_1^{(3)}},{\overline S_1^{(2)}}$         \\
            & $p_i \geq 0,\ i=1,2,3$            &       \\
            & $p_7 > 0\,,\;q+p_3(p_1+p_2)>0$          &         \\
 $S^{(4)}_2\equiv \tilde{S}^{(3)}_3$ & $ s_1=q=0=p_i\,, \; i=3,5,9,10\,; \; p_7>0\,$ & ${\overline S_2^{(3)}},{\overline S_1^{(2)}}$          \\
                 & $p_i \geq 0\,,\;i=1,2,4\,;\; p_4(p_1+p_2)+ p_8 > 0$                &                                                        \\
$S^{(4)}_3$ & $p_i = 0,\ i=4,6,7,8,9,10$                   & ${\overline S_1^{(3)}}$                                \\
            & $p_i >0,\ i=1,2,3\,; \; p_1 p_2 p_3 - p_5^2>0$&   \\
 $S^{(3)}_1 \equiv \tilde{S}^{(2)}_1$ & $p_5^2 - p_1p_2p_3 =  0=p_i ,\ i=4,6,7,8,9,10 $        & ${\overline S_i^{(1)}},$ $i=1,2,3$ \\
            & $p_i \geq 0, \ i=1,2,3\,; p_1 p_2 +p_3 (p_1+p_2)>0$          &           \\
$S^{(3)}_2 \equiv \tilde{S}^{(2)}_2$ & $s_1 = q = 0= p_i ,\ i=3,5,7,9,10$  & ${\overline S_i^{(1)}}, $ $i=1,2,4$  \\
            & $p_i \geq 0, \ i=1,2\,;\; p_1 p_2 +p_4 (p_1+p_2)>0 $         &          \\
 $S^{(2)}_1 \equiv \tilde{S}^{(2)}_3$ & $p_7 - p_1p_2 =  p_i =0 <p_1\;,p_2,\;\; i\ne 1,2,7$     & ${\overline S_1^{(1)}}\,,\;{\overline S_2^{(1)}}$   \\
 $S^{(2)}_2$ & $p_i = 0,\ i\ne 3,4 \,; \;0< p_3\;,p_4$                      & ${\overline S_3^{(1)}}\,,\;{\overline S_4^{(1)}}$   \\
$S^{(1)}_1\equiv \tilde{S}^{(1)}_1$ & $p_i = 0 < p_1,\; \ i\ne 1$            & $S^{(0)}$ \\
$S^{(1)}_2 \equiv \tilde{S}^{(1)}_2$ & $p_i = 0< p_2,\; \ i\ne 2$          & $S^{(0)}$ \\
$S^{(1)}_3$ & $p_i = 0 <p_3, \ i\ne 3$                       & $S^{(0)}$ \\
$S^{(1)}_4$ & $p_i = 0<p_4, \ i\ne 4$                    & $S^{(0)}$ \\
$S^{(0)}\equiv \tilde{S}^{(0)}$   & $p_i = 0,\; \;\;1\leq  i \leq 10$        &  \\
\hline
\end{tabular}
\end{center}

\end{table}

%%%%%%%%%%%%%%%%%%%%%%%%%%%%%%%%%%%%%

%%%%%%%%%%%%%%%%%%%%%%%%%%%%%%%%%%%%%%%%%%
\begin{table}
\caption{\label{T2b} \footnotesize Symmetries of
the strata $S$ of Model 3 and $\tilde{S}$ of Model 2.
The group
$\mathrm{U}_1^{\mathrm{e.m.}}$ is defined as in Table~(\ref{T1}),
and
$\alpha = \mathrm{e}^{\mathrm{i}\pi(T_3-Y/2)}$.
 Symmetries are specified by a {\em
representative element} of the conjugacy class of isotropy
subgroups. Finite groups are defined through their generators
between brackets.
For each stratum, a field configuration with the same symmetry is supplied {\em (typical point)}.
The fields are represented by their real components:
$\Phi_1^{\mathrm T}=(\phi_1 + \mathrm{i} \phi_2,\phi_3 + \mathrm{i} \phi_4)$,
$\Phi_2^{\mathrm T}=(\phi_5 + \mathrm{i} \phi_6,\phi_7 + \mathrm{i} \phi_8)$,
$F=\phi_9+\mathrm{i} \phi_{10}$
and $H=\phi_{11}$, where
 the $\phi_i$'s are generic non zero values.
}
\begin{center}
\begin{tabular}{lcl}
\hline
 Stratum   &  Symmetry         &   Typical point $\phi$ \\
 \hline
 $S^{(7)} \equiv \tilde{S}^{(4)} $  &  $\{\uno\}$ & $(\phi_1,0,\phi_3,\phi_4,0,0,\phi_7,0,\phi_9,\phi_{10},\phi_{11})$\\
 $S^{(5)}\equiv \tilde{S}^{(3)}_1$   & $\{\hat\iota\,K\}$  & $(\phi_1,0,\phi_3,0,0,0,0,\phi_8,\phi_9,0,\phi_{11})$ \\
 $S^{(4)}_1\equiv \tilde{S}^{(3)}_2$ &  $\{K\}$            & $(\phi_1,0,\phi_3,0,0,0,\phi_7,0,0,\phi_{10},0)$   \\
 $S^{(4)}_2\equiv \tilde{S}^{(3)}_3$ & U$_1^{\mathrm e.m.}$  & $(0,0,\phi_3,\phi_4,0,0,\phi_7,0,0,0,\phi_{11})$  \\
 $S^{(4)}_3$ & $\{\alpha\,\hat\iota\}$         & $(\phi_1,0,0,0,0,0,\phi_7,0,\phi_9,\phi_{10},0)$     \\
 $S^{(3)}_1 \equiv \tilde{S}^{(2)}_1$ &  $\{\alpha\,\hat\iota ,K\}$      & $(\phi_1,0,0,0,0,0,\phi_7,0,0,\phi_{10},0)$  \\
$S^{(3)}_2 \equiv \tilde{S}^{(2)}_2$  & U$_1^{\mathrm e.m.}\times\,\{\hat\iota\,K\}$ & $(0,0,\phi_3,0,0,0,0,\phi_8,0,0,\phi_{11})$ \\
 $S^{(2)}_1 \equiv \tilde{S}^{(2)}_3$ &  U$_1^{\mathrm e.m.}\times\{K\}$     & $(0,0,\phi_3,0,0,0,\phi_7,0,0,0,0)$   \\
$S^{(2)}_2$ &  SU$_2\times\{\hat\iota \,K\}$   & $(0,0,0,0,0,0,0,0,\phi_9,0,\phi_{11})$   \\
$S^{(1)}_1\equiv \tilde{S}^{(1)}_1$ &  U$_1^{\mathrm e.m.}\,\times\{\hat\iota, K\}$      & $(0,0,\phi_3,0,0,0,0,0,0,0,0)$ \\
$S^{(1)}_2 \equiv \tilde{S}^{(1)}_2$ &  U$_1^{\mathrm e.m.}\times \{e^{i\pi Y}\,\hat\iota, K\}$ & $(0,0,0,0,0,0,\phi_7,0,0,0,0)$ \\
$S^{(1)}_3$ & SU$_2\times\{\hat\iota \, K\,,  \mathrm{e}^{\mathrm{i} \pi Y/2}\,\hat \iota\}$    & $(0,0,0,0,0,0,0,0,\phi_9,0,0)$ \\
$S^{(1)}_4$ & $(\mathrm{SU}_2\times \mathrm{U}_1)/\integer_2\times\{\hat\iota \,K\}$  & $(0,0,0,0,0,0,0,0,0,0,\phi_{11})$ \\
$S^{(0)}\equiv \tilde{S}^{(0)}$   &  $(\mathrm{SU}_2\times \mathrm{U}_1)/\integer_2\,\times \{\hat \iota\,,\;K\}$       & $(0,0,0,0,0,0,0,0,0,0,0)$ \\
\hline
\end{tabular}
\end{center}

\end{table}

%%%%%%%%%%%%%%%%%%%%%%%%%%%%%%%%%%%%%

We shall consider two different dynamical versions of Model 2, a
complete non-renormalizable and an incomplete renormalizable one.

\subsection{Model $2_1$: A complete non-renormalizable version of Model 2}
For the reasons explained in the Introduction, let us now ignore
the renormalizability condition. Then, the simple
potential defined in \eref{V}, with the ${\tilde p}_i$'s specified
in \eref{IB2} replacing the $p_i$'s,
is already sufficient (despite its being not the
most general $G$-invariant eighth-degree polynomial), to make
observable all the allowed phases. It admits, in fact, a stable
minimum in each of the strata ${\tilde S}$ listed in
Tables~\ref{T2a} and \ref{T2b}, for suitable values of the $a_i$'s, as can be
easily realized, with the help of Figure~\ref{F1},
conveniently modifying
the  geometrical arguments exploited to determine the
observable phases of Model 1.
The transformation in Eq.~\eref{proinv} leads to a four dimensional
semialgebraic set which, contrary to $\Xi$, is not convex, but, fortunately,
like $\Xi$, has no
intruding cusps. In
particular, for $A\propto \uno$ let us denote by $\Omega$ the
 image, $\eta=p(\real^8)$, of the orbit space in the
space of the $\eta_i$'s. Then, if $\eta$ is within or near enough to
$\Omega$,
there is only one local minimum of the
potential (the absolute minimum) at the point $\tilde{p}$ of the orbit
space which is at the shortest distance from $\eta$.

The above statements have been checked analytically; in particular,
we have determined the superposition regions
in the $\eta$ parameter space, where two local minima of
potential \eref{V} can
co-exist in two different symmetry strata \cite{mat}.
Such a situation is quite important from a phenomenological point of
view, since it can lead, even at the classical level,
to first order phase transitions \cite{5}.

\subsection{Model $2_2$: An incomplete renormalizable version of Model 2}
Let us now, in the frame of symmetries of Model 2, chose the Higgs
potential as the most general, bounded below invariant polynomial
of {\em degree four:}

\begin{equation}
\widehat V(\tilde{p})= \sum_{i,j=1}^2 \,A_{ij}\,\tilde{p}_i\,\tilde{p}_j
+ \sum_{i=1}^4 \,a_i\,\tilde{p}_i,\label{V3}
\end{equation}
where all the parameters are real and $A_{12}=A_{21}$.

As stated in \cite{AS}, since there are no relations among the
basic invariants and the potential is linear in those of degree
four, its local minima can only sit on the boundary of the orbit
space, for general values of the $a_i$'s. A detailed calculation
shows that there can be stationary points of the potential in the
strata of dimension $\ge 3$ only if the $a_i$'s satisfy particular
conditions: $a_3=a_4=0$, $a_4=0$, $a_3=0$ and $a_3=a_4$,
respectively, for the strata $\tilde{S}^{(4)}$,
$\tilde{S}^{(3)}_1$, $\tilde{S}^{(3)}_2$ and $\tilde{S}^{(3)}_3$.
These conditions reduce to zero the measures of the regions of
stability of the corresponding phases, in the space of the
parameters $a_i$. So, there will not be stable phases associated
to the strata of dimension $\ge 3$. As a consequence, it is
impossible to generate spontaneous $CP$ violation in the model.
The general problem of spontaneous $CP$ breaking in two-Higgs
doublet models will be faced in a forthcoming paper \cite{5}.

We can conclude that Model $2_2$ is renormalizable, but it is  {\em
incomplete}. In tree-level approximation, the phases associated to the
strata of dimension $\geq 3$ have zero probability to be observed
in nature.

\section{Model 3}
The comparison of models $2_1$ and $2_2$ suggests a way to make
observable all the phases allowed by the symmetry of Model 2,
without changing the symmetry group $G= (\mathrm{SU}_2\times \mathrm{U}_1)/\integer_2
\,\times \{\hat \iota\,,\, K\}$. If
some of the basic SU$_2$ invariant polynomials which have been
used to build up the basic invariants of Model 2 are considered as
independent singlet scalar fields, the potential of Model $2_1$, written
in terms of the new fields, can take on a renormalizable form.
Here, the new SU$_2$ singlet scalar fields are added only as a
technical trick, but they could be thought of as describing bound
states of the original scalar doublets. The introduction of
additional scalar fields in a model modifies the symmetry and
might enlarge the number of allowed phases. However, one can
get rid of possible new phases, by making them dynamically
unattainable.

Let us just analyze one of these possibilities, that we shall call
Model 3, which is obtained from Model 2 by adding a complex,
hypercharge 2 singlet $F$ and a real, hypercharge 0 singlet $H$
with transformation rules $(F,H)\rightarrow (-F,-H)$ and,
respectively, $(F,H)\rightarrow (-F^*,-H)$ under transformations
induced by $\hat\iota$ and $K$.

The following set of invariants yields a MIB in the present case:
\eqll{\begin{array}{l}
p_1=\tilde{p}_1,\ \ p_2=\tilde{p}_2,\ \ p_3=|F|^2,\ \ p_4=H^2,\ \
p_5=\mathrm{Re}[\Phi_2^{\mathrm T}\sigma_2\Phi_1\,F^*],\ \\
 p_6=\mathrm{Im}\left[\Phi_2^\dagger\Phi_1\right]H,\;
 p_7 = \tilde{p}_3,\; p_8 = \tilde{p}_4,\;
 p_9 = H\,\mathrm{Re}\left[\Phi_2^\dagger\Phi_1\right]\mathrm{Im}[\Phi_2^{\mathrm
T}\sigma_2\Phi_1\,F^*],\\
  p_{10}= \mathrm{Re}\left[\Phi_2^\dagger\Phi_1\right]
  \mathrm{Im}\left[\Phi_2^\dagger\Phi_1\right]
  \mathrm{Im}[\Phi_2^{\mathrm
T}\sigma_2\Phi_1\,F^*],
\end{array}}{IB3}
where $\tilde{p}_j$, $j=1, \dots ,4$ are defined in (\ref{IB2}).

The elements of the MIB have degrees $\{2,2,2,2,3,3,4,4,6,7\}$ and
are related by the syzygies $s_i=0$, $1\leq i \leq 6$:
\eqll{\begin{array}{rcl} s_1 &=& p_6^2 - p_4\, p_8,\ s_2 = p_9^2 -
p_4\, p_7\left(p_3 q - p_5^2\right), \
s_3 = p_{10}^2 - p_7\, p_8\left(p_3 q - p_5^2\right),\\
s_4 &=&  p_6\, p_9 - p_4\, p_{10},\ s_5 = p_6\, p_{10} - p_8\,
p_9,\ s_6 = p_9\, p_{10}- p_6\, p_7\left(p_3 q - p_5^2\right),
\end{array}}{Sy}
 where $ q=p_1\, p_2 - p_7 - p_8$.
Only three of the syzygies are independent. Therefore, the orbit
space is a semialgebraic subset of the seven dimensional algebraic
variety defined by the set of equations $s_i=0$, $i=1,\dots ,6$.
As usual, the orbit space and its stratification can be determined
from rank and positivity conditions of the $\widehat P(p)$-matrix
associated to the MIB defined in (\ref{IB3}). The results are
reported in  Tables~\ref{T2a} and \ref{T2b}. As expected, four new phases
({\em i.e.:} $S^{(4)}_3$, $S^{(2)}_2$, $S^{(1)}_3$, $S^{(1)}_4$),
that were not allowed by the symmetry of Model 2, are now allowed.

The most general invariant polynomial of  degree four in the
scalar fields of the model can be written in terms of the
following polynomial $\widehat V(p)$ in the $p_i$'s with degree
$\le 4$:

\eqll{\widehat V(p)=\sum_{i=1}^8\,a_i\,p_i + \sum_{i,j=1}^4\,A_{ij}\,p_i\,p_j.}{V3'}

The conditions for a stationary point of $\widehat V(p)$ in a
given stratum  are obtained from equation
\eref{ext} and the explicit form of the $f_\alpha(p)$ can be read
from Table~\ref{T2a}.

The high dimensionality of the orbit space prevents, in this case,
a simple geometric determination of the conditions assuring the
existence of a stable local minimum on a given stratum
 and a complete analytic solution of these conditions
is impossible, since high degree polynomial equations are
involved. Despite this, using convenient majorisations, we have
been able to prove \cite{mat} that all the phases allowed by the symmetry of
Model 2 (which are, obviously, also allowed by the symmetry of
Model 3) are observable in Model 3. The result holds in the
following strong sense: For each phase of symmetry $(H)$ allowed
by the symmetry of Model 2, we have analytically determined an
8-dimensional open semialgebraic set ${R}_H$ in the space of the
coefficients $a=(a_1,\dots ,a_8)$ and an open neighborhood ${I}_H$
of the four dimensional unit matrix, such that, for all $a\in
{R}_H$ and $A\in{I}_H$, the potential $\widehat V(p(\phi))$,
defined through (\ref{V3'}), has a stable absolute minimum in the
stratum of Model 3 with symmetry $(H)$. Moreover, for $a\in {R}_H$
and $A\in{I}_H$, the potential has no local minimum in the strata
of Model 3 corresponding to phases not allowed in Model 2.  In
particular:

\begin{enumerate}
\item All the allowed phases of Model 3
turn out to be observable with the potential defined in
(\ref{V3'}), but for the phase corresponding to the stratum
$S^{(4)}_3$ (a local minimum lies in this stratum only if $a_5=0$,
which means that the minimum is unstable), so that Model 3, which
is renormalizable, is not far from being also complete;

\item For $A=\uno$, a stable local minimum can be found only
in the strata of Model 3 corresponding to allowed phases of Model
2, provided that $a_1 + a_2 < 0$ and $a_3\,a_4 > 0$.
\end{enumerate}
A complete specification of the sets ${R}_H$ and ${I}_H$ would be too long to
be given in this Letter and will be reported elsewhere \cite{mat}.

Let us conclude by stressing that Model 3 could be relevant in the study
of electro-weak baryogenesis: $CP$ violation is achieved in phase
$S^{(4)}_2$, so it is interesting to examine the possibility of first
order phase transitions to more symmetrical phases \cite{5}.


\begin{thebibliography}{00}
\bibitem{isos}
We are not interested in possible isostructural phase transitions
taking place when a continuous variation of the $a$'s induces a
discontinous displacement (a jump) of the location of the minimum to a
different point of the same stratum.
\bibitem{nota}
The invariant
$p_1(x)=\sum_{i=1}^n\,x_i^2$ can always be included in a MIB and,
in this case,
the image in $\real^q$ of the unit sphere of $\real^n$ is the hyperplane
of equation $p_1=1$.
\end{thebibliography}
\end{document}